\documentstyle[preprint,prd,aps,eqsecnum]{revtex}
\tighten
\begin{document}
\draft
\preprint{gr-qc/9410039\qquad UCSBTH-nn-nn}
\begin{title}
	{Dynamical Systems Treatment of Scalar Field Cosmologies}
\end{title}
\author			{Shawn J.~Kolitch\\
		{\tt kolitch@nsfitp.itp.ucsb.edu}}
\address{Department of Physics\\University of California\\
		Santa Barbara, CA 93106-9530}

\author			{Brett Hall\\
		{\tt brett@vorpal.physics.ucsb.edu}}
\address{Department of Physics\\University of California\\
		Santa Barbara, CA 93106-9530}

\date{October  , 1994}

\maketitle
\begin{abstract}
The conformal equivalence of some cosmological models in Brans-Dicke
theory to general relativistic cosmologies with a scalar field is
discussed.  In the case of radiation-dominated universes, it is shown
that the presence of the scalar field has a negligible impact upon the
evolution of the models in the Einstein frame.  It is also shown that
power-law inflation in general relativity, which is conformal to
``extended'' power-law inflation in Brans-Dicke theory, is not a unique
attractor for expanding closed universes, but rather that the occurence
of inflation depends upon the initial kinetic energy of the scalar field.
\end{abstract}

\def\fig#1{Fig.~{#1}}			
\def\Equation#1{Equation~\(#1)}		
\def\Equations#1{Equations~\(#1)}	
\def\Eq#1{Eq.~\(#1)}			
\def\Eqs#1{Eqs.~\(#1)}			
\let\eq=\Eq\let\eqs=\Eqs		
\def\(#1){(\call{#1})}
\def\call#1{{#1}}
\def\square{\kern1pt\vbox{\hrule height 1.2pt\hbox{\vrule width 1.2pt\hskip 3pt
   \vbox{\vskip 6pt}\hskip 3pt\vrule width 0.6pt}\hrule height 0.6pt}\kern1pt}

\section{Introduction}
\label{intro}
In inflationary universe scenarios, it is usually assumed that general
relativity (GR) is the correct theory of gravitation, and the matter is
generally taken to be a homogeneous scalar field $\phi$, with a potential
$V(\phi)$ acting as the vacuum energy to drive the accelerated expansion.
The gravitational action and the form of this potential should, in principle,
be motivated by the fundamental physics which governs the very early
universe, and which would also give rise to the scalar field itself.
However, as such a theory is not yet definitely known, it is of
interest to study a wide variety of scenarios.  In GR, for example,
if the potential is simply a constant, $V(\phi) = V_0$, then
the spacetime is deSitter and the expansion is exponential.  If the
potential is exponential, $V(\phi) = V_0 e^{-\lambda\phi}$, then there
is a power-law inflationary solution\cite{p-law}.
Models of ``extended'' inflation\cite{La}, on the other hand, choose
Brans-Dicke (BD) theory\cite{BD} as the correct theory of gravity,
and in this case the choice of a constant vacuum energy yields a
power-law solution directly\cite{BDinf}, whereas exponential expansion
may only be obtained if a cosmological constant is explicity inserted
into the field equations\cite{Romero,Barrow90,Stlam}, in which case it
cannot be interpreted as the vacuum energy of any physical matter field.
It is well known that the former case is conformally equivalent to
power-law inflation in general relativity\cite{conformaleq}.

Recently, it was shown that the field equations for cosmology in Brans-Dicke
theory may be reduced to a two-dimensional dynamical system for any
reasonable perfect fluid matter source\cite{Stlam,Kolitch94,Wands}.  This
is possible for all values of spatial curvature if there is no cosmological
constant, and also for spatially flat models with a nonzero cosmological
constant.  Here we point out that the conformal transformation between
the Brans-Dicke frame and the Einstein frame preserves this dynamical
system treatment in some cases, and we use this fact in order to make
comments about some general relativistic cosmological models, based upon
results which have previously been obtained in the Brans-Dicke frame.

The paper is organized as follows.  In Sec.~II, we discuss the conformal
equivalence of Brans-Dicke theory to general relativity with a scalar field.
In Sec.~III, we summarize our dynamical system formalism for BD cosmologies
and state its relevance to the Einstein frame, and in Sec.~IV we apply the
formalism to the models of interest.  Sec.~V presents conclusions.

\section{The Brans-Dicke and Einstein Frames}

\subsection{Conformal Equivalence}
Including a possible cosmological constant, Brans-Dicke theory may be
formulated in terms of the action
\begin{equation}
	S_{BD} = {1\over {16\pi}}\int d^4x \sqrt{-g} \left[\phi (R - 2\Lambda)
		- \omega {{g_{\mu\nu}\phi^{,\mu}
		\phi^{,\nu}}\over {\phi}} - 16\pi
		{\cal \char'114}_m \right].
\label{BDaction}
\end{equation}
As Dicke himself pointed out in a companion to his seminal paper with
Brans\cite{Dicke}, one can reformulate this theory as one in which the
Einstein equations are formally satisfied via the conformal transformation
\begin{equation}
	\tilde g_{\mu\nu} = G\phi g_{\mu\nu},
\label{confxf}
\end{equation}
where $G$ is the value Newton's gravitational constant observed today.
If one makes the further field redefinition
\begin{equation}
	G\phi = \exp \Biggl(\psi\sqrt{{8\pi G}\over {\omega + 3/2}}\; \Biggr),
\label{redef}
\end{equation}
one arrives at the new action
\begin{equation}
	\tilde S_{BD} = {1\over {16\pi G}}\int d^4x \sqrt{- \tilde g}
		\left[(\tilde R - 2\Lambda) - 16\pi G
		(\tilde{\cal \char'114}_m + {1\over 2}
		\tilde g_{\mu\nu} \psi^{,\mu}\psi^{,\nu})
		\right],
\end{equation}
where
\begin{equation}
	\tilde{\cal \char'114}_m = \exp \left(-2\psi\sqrt{{8\pi G}\over
		{\omega + 3/2}} \; \right){\cal \char'114}_m.
\end{equation}
This is the action for general relativity with a massless scalar field
which is exponentially coupled to the other matter.

\subsection{Field Equations}
The field equations for Brans-Dicke theory, obtained by varying
Eq.~\(\ref{BDaction}), are
\begin{equation}
	2\omega \phi^{-1} \square \phi - \omega {{\phi^{,\mu}
	\phi_{,\mu}} \over {\phi^2}} + R - 2\Lambda = 0
\label{BDphieq}
\end{equation}
and
\begin{equation}
	R_{\mu \nu} - {1\over 2}g_{\mu \nu}R + g_{\mu \nu}\Lambda
		= 8\pi \phi^{-1} T_{\mu \nu} +
		\left({\omega \over \phi^2} \right)(\phi_{,\mu} \phi_{,\nu}
 		- {1\over 2}g_{\mu \nu} \phi^{,\rho} \phi_{,\rho})
		+ \phi^{-1}(\phi_{,\mu {;}\nu} - g_{\mu \nu}
		\square \phi),
\label{BDgeq}
\end{equation}
where
\begin{equation}
	\square \phi \equiv {1 \over \sqrt{-g}}{\partial \over
		{\partial x_\mu}}\left(\sqrt{-g}{{\partial \phi}
		\over {\partial x^\mu}}\right).
\label{boxdef}
\end{equation}
Taking the trace of Eq.~\(\ref{BDgeq}) and combining the result with
Eq.~\(\ref{BDphieq}), one finds
\begin{equation}
	\square \phi = \left({8\pi} \over {3 + 2\omega} \right)
		\left(T^\mu {}_\mu - {{\Lambda \phi} \over 4\pi}\right).
\label{BDboxeq}
\end{equation}
One generally takes Eqs.~\(\ref{BDgeq}) and (\ref{BDboxeq}) as the
independent field equations.

In the Einstein frame, one of course recovers the usual Einstein equations,
\begin{equation}
	\tilde R_{\mu\nu} - {1\over 2}\tilde g_{\mu\nu}\tilde R
		+ \tilde g_{\mu\nu}\Lambda = 8\pi \tilde T_{\mu\nu},
\end{equation}
where now
\begin{eqnarray}
	\tilde T_{\mu\nu} =&& \tilde T_{\mu\nu}^{\psi}
		+ \tilde T_{\mu\nu}^m	\nonumber			\\
	=&& \psi_{,\mu}\psi_{,\nu} - {1\over 2}\tilde g_{\mu\nu}
		\tilde g_{\rho\sigma}\psi^{,\rho}\psi^{,\sigma}
		+ \tilde T_{\mu\nu}^m.
\end{eqnarray}
It then follows from the Bianchi identities that $\tilde T^{\mu\nu}{}_{;\nu}
= 0$, from which one sees that the ordinary matter will obey the proper
conservation equation, ${(\tilde T^{\mu\nu}{}_{;\nu})}^m = 0$, only if
the corresponding equation of motion for the scalar field is satisfied,
{\it i.e.,\ } only if ${(\tilde T^{\mu\nu}{}_{;\nu})}^{\psi} = 0$, which
implies that ${ \tilde {\square} } \psi = 0$.

Under the double transformation given by Eqs.~\(\ref{confxf}) and
(\ref{redef}), one finds that Eq.~\(\ref{BDboxeq}) becomes
\begin{equation}
	e^{2c_1\psi}{ \tilde {\square} } \psi =
		{c_1 \over 2}\left[T^{\mu}{}_{\mu} - {\Lambda \over {4\pi G}}
		e^{c_1 \psi} \right],
\label{boxxf}
\end{equation}
where $c_1 = \sqrt{8\pi G / (\omega + 3/2)}$.
Hence we see that two conditions must be met if results from the Brans-Dicke
frame are to be applied to realistic Einstein universes with a scalar field:
the stress-energy of the ordinary matter (neglecting the scalar field)
must be traceless, and the cosmological constant must vanish.
Cases of interest which satisfy these conditions are true-vacuum or
radiation-filled universes, and power-law inflationary universes.

\section{Dynamical System Formalism}
In \cite{Kolitch94}, it was shown that the field equations for perfect
fluid Brans-Dicke cosmologies with vanishing cosmological constant may
be reduced to a planar dynamical system, and this system was analyzed
qualitatively for several cases of interest.  Here we will simply outline
the method.  First we switch to conformal time $d\tau = dt / a(t)$, and
define the new variables
\begin{eqnarray}
	\beta \equiv&& \left({a' \over a} + {\phi ' \over {2\phi}} \right),
		\label{betadef}						\\
	\sigma \equiv&& A {\phi' \over \phi},	\label{sigdef}
\end{eqnarray}
where primes represent derivatives with respect to conformal time, and
\begin{equation}
	A \equiv \left({{2\omega+3}\over 12}\right).
\label{Adef}
\end{equation}
Next, we parametrize the equation of state by writing
\begin{equation}
	p = (\gamma - 1) \rho.
\label{pdef}
\end{equation}
Writing out Eqs.~(\ref{BDgeq}) and (\ref{BDboxeq}) for a FRW metric,
assuming that the matter satisfies the usual conservation equation, and
using Eqs.~(\ref{betadef})--(\ref{pdef}), one can derive the dynamical system
\begin{eqnarray}
	\sigma' =&& (1-3\gamma /4)(\beta^2 + k - \sigma^2 /A) - 2\beta \sigma,
		\label{DS1}						\\
	\beta' =&& (1-3\gamma /2)(\beta^2 + k) - (3-3\gamma /2)\sigma^2 /A.
		\label{DS2}
\end{eqnarray}
In the special case of the true vacuum $(\rho = 0)$, the system reduces to
\begin{eqnarray}
	\sigma' =&& -2\beta\sigma,					\\
	\beta' =&& -2\sigma^2 / A,
\end{eqnarray}
which are just Eqs.~(\ref{DS1}) and (\ref{DS2}) subject to the ``vacuum
constraint'' $\beta^2 + k - \sigma^2 /A = 0$ (equivalent to the
zeroth component of Eq.~(\ref{BDgeq}) with $\rho = 0$).  These systems may
now be analyzed using the standard methods of dynamical systems
theory\cite{DSref}.

One can easily verify that under the transformation to the Einstein frame
specified by Eqs.~(\ref{confxf}) and (\ref{redef}), the dynamical system
(\ref{DS1}, \ref{DS2}) is unaltered, provided that we make the additional
redefinitions $\beta = b'/b$ and $\sigma = Ac_1 \psi'$, where $b$
is the metric in the Einstein frame and the constants $A$ and $c_1$ were
defined in Eqs.~(\ref{boxxf}) and (\ref{Adef}).  Note, however, that there
will be non-geodesic motion of the cosmological fluid embedded in these
equations unless the stress-energy of the ordinary matter is traceless,
for reasons which were explained in Sec. II.

\section{Results}

Although one can obtain the significant features of the solutions to
Eqs.~(\ref{DS1}) and (\ref{DS2}) analytically, a useful technique is to
numerically integrate the system, and then to plot the solution curves
(trajectories) in the $\beta$--$\sigma$ plane.  Figs.~1 and 2 show the
results of this procedure for universes dominated by radiation and
false-vacuum energy, respectively.  The shaded regions require negative
energy density and so are disallowed on physical grounds.  The boundaries
of these regions represent true vacuum solutions; they are just those
trajectories which satisfy the ``vacuum constraint'' mentioned previously.
In the case of $k =0$, these solutions were found analytically by O'Hanlon
and Tupper\cite{TV}.  The dark solid line represents the static condition
$da/dt = 0$ in the Brans-Dicke frame; the dark dashed line represents the
analagous condition $db/dt = 0$ in the Einstein frame.  Note that, in general,
there is a family of solutions for any matter source and spatial curvature,
with the models parametrized by the value of $\sigma$ at a given $\beta$.
Also note that when $\omega < 0$, nonsingular (for $k = 0,-1$) and
perpetually expanding (for $k = +1$) radiation models are possible in
the Brans-Dicke frame, due to the domination of the scalar field in the
dynamics.  The presence of these models corresponds to the rotation of
the line $da/dt = 0$ into
the physical regime of the $\beta$--$\sigma$ plane.  In the Einstein frame,
where the scalar is not coupled directly to the metric, this behavior
is not possible unless the matter is inflationary, {\it i.e.,\ } unless
the strong energy condition $p > -\rho /3$ is violated,
consistent with the fact that such models are precluded in GR by
singularity theorems\cite{Singthm}.

Now one can read the possible behavior of the models in either conformal
frame directly from the figures.  Fig.~1a represents the spatially flat,
radiation-filled FRW models.  One sees that all of the initially expanding
models approach the equilibrium point $(\beta_0, \sigma_0) = (0,0)$, which
represents the usual quasi-static endpoint of a flat-space FRW universe.
In the BD frame, we have chosen $\omega = -1/2$, so that nonsingular models
which pass smoothly from contraction to expansion are present.  This
possibility does not exist in the Einstein frame, where all of the
contracting models continue to contract until a singularity is reached.
In fact, as must be the case, there is no $\omega$--dependence at all
among the radiation models in the Einstein frame, as one can see by
substituting $\gamma = 4/3$ and $\sigma = Ac_1 \psi' \sim \psi'
\sqrt{2\omega + 3}$ into Eqs.~(\ref{DS1}) and (\ref{DS2}).  Negative curvature
models are represented in Fig.~1b, and the results are identical to those for
flat-space models, except that the late-time attractor for the expanding
models represents linear expansion, as one would expect for open universes.
If the curvature is positive as represented by Fig.~1c, then in the Einstein
frame all of the models recollapse as one would expect, although in the
BD frame perpetual expansion is possible for certain initial conditions.
One concludes that for the radiation-dominated universes, the presence
of the scalar field has a negligible effect upon the dynamics of the models
in the Einstein frame.

On the other hand, consider a constant vacuum energy in the BD frame,
such as would be supplied by an inflaton field which was trapped in a
false minimum of its potential, and which produces power-law inflation
in both conformal frames.  In the Einstein frame, one is considering
a scalar field with exponential potential $V(\psi) \sim e^{-\lambda\psi}$,
where $\lambda$ is related to $\omega$ by $\lambda = [32\pi G /
(\omega + 3/2)]^{1/2} = 2c_1$\cite{conformaleq}.  For many expanding
models there is an attractor,
which is the power-law solution first found by Mathiazhagan and
Johri\cite{BDinf}, and later used by La and Steinhardt in their original
model of extended inflation\cite{La}.  The solution is $a(t) \sim
t^{\omega + 1/2}$ in the BD frame; in the Einstein frame it is $b(t) \sim
t^{\omega / 2 + 3/4}$.  Clearly the value $\omega = 1/2$ seperates models
which can inflate from those which cannot.  Consider then only those models
which fall in the inflationary part of the parameter space, {\it i.e.,\ }
consider $\omega > 1/2$ or, equivalently, $\lambda < (16\pi G)^{1/2}$.
In Figs.~2a and 2b, we have chosen $\omega = 10.5$, and one sees that
the zero- and negative-curvature models all inflate eventually.
However, it is apparent from Fig.~2c that models with positive curvature do
not inflate unconditionally; rather, for some initial conditions recollapse
occurs.  The same statement therefore holds in both conformal frames:  if
the rate of change of the scalar field is too high relative to the expansion
rate of the universe, recollapse rather than inflation will occur.  In the
case of power-law inflation in GR, this was first shown by Halliwell using
a slightly different formalism\cite{Halliwell}.

\section{Conclusions}
We have seen that the conformal equivalence of BD theory to GR
with a scalar field preserves a dynamical systems treatment of
cosmological models in some cases of interest, including radiation-filled
universes and power-law inflationary models.  In the case of radiation,
the presence of the scalar field has a negligible impact upon the
late-time behavior of the models.  This is not surprising, as the
matter in this case is completely decoupled from scalar field in both
conformal frames.  In the case of power-law inflation, one finds that
if the scalar field dominates the dynamics, some positively-curved
models will recollapse rather than inflate.  Hence one cannot assume,
either in extended inflationary scenarios in Brans-Dicke theory, or in
power-law inflationary scenarios in general relativity, that accelerated
expansion will actually occur.  In particular, positively curved
universes where the energy of the scalar field is high relative to the
expansion rate may not inflate.  Hence the initial conditions
of the inflationary epoch must be taken into consideration.  One
can speculate that similar statements may hold in other inflationary
scenarios, such as ``hyperextended inflation'' in Brans-Dicke
theory\cite{S-A}, or models in GR where the scalar field has a
polynomial potential.

\section{Acknowledgements}
S.K. was supported in part by the National Science Foundation under
Grants No.~PHY89-04035 and PHY90-08502.  B.H. was supported by a SURF
fellowship from the College of Creative Studies at UC Santa Barbara.

\gdef\journal#1, #2, #3, 1#4#5#6{		
    {\sl #1~}{\bf #2}, #3 (1#4#5#6)}		
\def\pr{\journal Phys. Rev., }
\def\pra{\journal Phys. Rev. A, }
\def\prb{\journal Phys. Rev. B, }
\def\prc{\journal Phys. Rev. C, }
\def\prd{\journal Phys. Rev. D, }
\def\prl{\journal Phys. Rev. Lett., }
\def\jmp{\journal J. Math. Phys., }
\def\rmp{\journal Rev. Mod. Phys., }
\def\cmp{\journal Comm. Math. Phys., }
\def\np{\journal Nucl. Phys., }
\def\pl{\journal Phys. Lett., }
\def\apj{\journal Astrophys. Jour., }
\def\apjl{\journal Astrophys. Jour. Lett., }
\def\annp{\journal Ann. Phys. (N.Y.), }

\begin{figure}
\caption{The evolution of radiation-filled universes in Brans-Dicke
theory with $\omega = -1/2$, and in general relativity with a massless
scalar field.  The shaded regions require $\rho < 0$ and so are disallowed
physically.  The dark solid line signifies $da/dt = 0$ in the BD frame;
the dark dashed line signifies $db/dt = 0$ in the Einstein frame.  In each
frame the trajectories to the right of the static line represent expanding
models, and those to its left represent contracting models.  Hence
nonsingular universes are possible only in the BD frame.  (a) Models
with vanishing spatial curvature ($k=0$).  All universes which start with
a big bang asymptotically approach quasistatic equilibrium at $(\beta_0,
\sigma_0) = (0,0)$.  (b) Negative-curvature models ($k = -1$).  The
endpoint for expanding models now represents linear expansion.  (c)
Positive-curvature models.  All models recollapse in the Einstein frame,
although some expand perpetually in the BD frame.
\label{fig1}}
\end{figure}

\begin{figure}
\caption{Models dominated by false-vacuum energy in the BD frame with
$\omega = 10.5$, corresponding to an exponential potential for the scalar
field in the Einstein frame with $\lambda = (8\pi G / 3)^{1/2}$.  (a) and (b)
represent flat-space and negative-curvature models, respectively, all of
which approach power-law inflation at late times.  (c) Positive-curvature
models.  Some models are possible which start from a big bang, but then
recollapse rather than inflate, due to the domination of the scalar field
in the dynamics.
\label{fig2}}
\end{figure}

\end{document}